\documentclass[a4paper,11pt]{article}
\pdfoutput=1 

\usepackage{jheppub} 

\usepackage[T1]{fontenc} 

\usepackage{amstext}
\usepackage{amsmath}
\usepackage{amssymb}
\usepackage[english]{babel}
\usepackage{graphicx}
\usepackage{rotating}
\usepackage{xcolor}
\usepackage{multirow}
\usepackage{url}
\usepackage{hyperref}

\title{\boldmath FACET: A new long-lived particle detector in the very forward region of the CMS experiment}

\author[a,1]{S.~Cerci,\note{Also at the Istanbul University.}}
\author[a,1]{D.~Sunar Cerci,}
\author[b]{D.~Lazic,}
\author[c,2]{G.~Landsberg,\note{Corresponding author.}}
\author[d]{F.~Cerutti,}
\author[d]{M.~Sabat\'e-Gilarte,}
\author[e,2]{M.G.~Albrow,}
\author[e]{J.~Berryhill,}
\author[e]{D.R.~Green,}
\author[e]{J.~Hirschauer,}
\author[f]{S.~Kulkarni,}
\author[g]{J.E.~Br\"ucken,} 
\author[h]{L.~Emediato,} 
\author[h]{A.~Mestvirishvili,}
\author[h]{J.~Nachtman,}
\author[h]{Y.~Onel,}
\author[h]{A.~Penzo,}
\author[i]{O.~Aydilek,} 
\author[i]{B.~Hacisahinoglu,}
\author[i,2]{S.~Ozkorucuklu,}
\author[i]{H.~Sert,}
\author[i]{C.~Simsek,}
\author[i]{C.~Zorbilmez,}
\author[j,1]{I.~Hos,}
\author[k]{N.~Hadley,}
\author[k]{A.~Skuja,}
\author[l]{M.~Du,}
\author[l]{R.~Fang,}
\author[l]{Z.~Liu,}
\author[m,1]{B.~Isildak}
\author[n,o]{and V.Q.~Tran}
\emailAdd{Salim.Cerci@cern.ch}
\emailAdd{Deniz.Sunar.Cerci@cern.ch}
\emailAdd{Dragoslav.Lazic@cern.ch}
\emailAdd{Greg.Landsberg@cern.ch}
\emailAdd{Francesco.Cerutti@cern.ch}
\emailAdd{Marta.Sabate.Gilarte@cern.ch}
\emailAdd{albrow@fnal.gov}
\emailAdd{Jeffrey.Berryhill@cern.ch}
\emailAdd{dgreen@fnal.gov}
\emailAdd{jhirsch@fnal.gov}
\emailAdd{suchita.kulkarni@uni-graz.at}
\emailAdd{jens.brucken@helsinki.fi}
\emailAdd{lregisem@cern.ch}
\emailAdd{Alexi.Mestvirishvili@cern.ch}
\emailAdd{Jane.Nachtman@cern.ch}
\emailAdd{yasar-onel@uiowa.edu}
\emailAdd{Aldo.Penzo@cern.ch}
\emailAdd{Orhan.Aydilek@cern.ch}
\emailAdd{Burak.Hacisahinoglu@cern.ch}
\emailAdd{Suat.Ozkorucuklu@cern.ch}
\emailAdd{Hale.Sert@cern.ch}
\emailAdd{Cagdas.Simsek@cern.ch}
\emailAdd{Caglar.Zorbilmez@cern.ch}
\emailAdd{Ilknur.Hos@cern.ch}
\emailAdd{hadley@umd.edu}
\emailAdd{skuja@umd.edu}
\emailAdd{mg1722004@smail.nju.edu.cn}
\emailAdd{141150012@smail.nju.edu.cn}
\emailAdd{zuoweiliu@nju.edu.cn}
\emailAdd{Bora.Isildak@cern.ch}
\emailAdd{vqtran@sjtu.edu.cn}
\affiliation[a]{Department of Physics, Adiyaman University, 02040, Adiyaman, Turkey}
\affiliation[b]{Physics Department, Boston University, 590 Commonwealth Ave, Boston, MA 02215, U.S.A.}
\affiliation[c]{Department of Physics, Brown University, 182 Hope St, Providence, RI 02912, U.S.A.}
\affiliation[d]{CERN, 1211 Geneva 23, Switzerland}
\affiliation[e]{Fermi National Accelerator Laboratory, PO Box 500, Batavia, IL 60510, U.S.A.}
\affiliation[f]{Institute of Physics, NAWI Graz, University of Graz, Universit\"atsplatz 5, A-8010 Graz, Austria}
\affiliation[g]{Helsinki Institute of Physics, University of Helsinki, Gustaf H\"allstr\"omin katu 2, 00560 Helsinki, Finland}
\affiliation[h]{Department of Physics and Astronomy, University of Iowa, 203 Van Allen Hall, Iowa City, IA 52242, U.S.A.}
\affiliation[i]{Physics Department, Istanbul University, Vezneciler Caddesi, 34134, Istanbul, Turkey}
\affiliation[j]{Department of Engineering Sciences, Istanbul University-Cerrahpasa, 34320 Avcilar, Istanbul, Turkey}
\affiliation[k]{Department of Physics, University of Maryland, College Park, MD 20742, U.S.A.}
\affiliation[l]{Department of Physics, Nanjing University, Nanjing 210093, China}
\affiliation[m]{Department of Natural and Mathematical Sciences, Ozyegin University, Orman Sk 13, 34794, Istanbul, Turkey}
\affiliation[n]{Tsung Dao Lee Institute, Shanghai Jiao Tong University, Shanghai 200240, China}
\affiliation[o]{Faculty of Fundamental Sciences, PHENIKAA University, Yen Nghia, Ha Dong, Hanoi 12116, Vietnam}

\abstract{We describe a proposal to add a set of very forward detectors to the CMS experiment for the high-luminosity era of the Large Hadron Collider to 
search for beyond the standard model long-lived particles, 
such as dark photons, heavy neutral  leptons, axion-like particles, and dark Higgs bosons.  The proposed subsystem is called \textbf{FACET} 
for \textbf{F}orward-\textbf{A}perture \textbf{C}MS \textbf{E}x\textbf{T}ension, 
and will be sensitive
to any particles that can penetrate at least 50~m of magnetized iron and decay 
in an 18~m long, 1~m diameter vacuum pipe. 
The decay products will be measured in detectors using identical technology to the planned CMS Phase-2 upgrade.}

\arxivnumber{2201.00019}

\begin{document}                                                            
\maketitle

\section{Introduction}
  
The existence of long-lived particles (LLPs), i.e., particles with the proper lifetime $c\tau$ in a macroscopic range, is predicted in many models of physics beyond the standard model (BSM). Generally, LLPs are naturally expected in models with small mass splittings between the adjacent states (e.g., in supersymmetric models) or with suppressed couplings to standard model (SM) particles. In particular, LLPs are often present in particle dark matter (DM) models, where they serve as portals between the DM and SM particles. The existence of DM is well established from astronomical observations and cosmology.
It is generally assumed that DM consists of BSM particles . While searches for such
particles in the TeV mass range continue at the CERN Large Hadron Collider (LHC), the possibility that new particles may be relatively light ($\lesssim$50~GeV),
and yet have escaped detection so far because of very weak coupling to SM particles, is receiving considerable attention, as discussed in, e.g., refs.~\protect\cite{Essig:2013lka,Alimena:2019zri,Agrawal:2021dbo}. There are many possible so-called portals;
these are neutral particles that couple weakly to the SM and also to DM particles
(but being unstable are not themselves DM candidates), such as dark photons~\cite{Okun:1982xi,Caputo:2021eaa,Araki:2020wkq}, 
heavy neutral leptons~\cite{Cottin:2021lrq,SHiP:2018xqw}, axion-like particles~\cite{Feng:2018pew,Bauer:2018uxu,dEnterria:2021ljz},
and scalars or dark Higgs particles~\cite{Duerr:2017uap,Feng:2017vli}. 

The proposed new CMS subsystem, FACET ({\bf F}orward-{\bf A}perture {\bf C}MS {\bf E}x{\bf T}ension), can search for many such 
LLPs (called here $X^0$), depending only on their forward production cross section, momentum, mass $m_{X^0}$, and proper lifetime $c\tau$. We will show that FACET covers a region of parameter space not accessible to other experiments, neither existing nor proposed. 

In what follows, we will use the CMS detector right-handed coordinate system~\cite{CMS:2008xjf}, with the origin at the CMS interaction point (IP5), with the $z$ axis pointing in the counterclockwise proton beam direction, the $y$ axis pointing upward, and the $x$ axis pointing toward the center of the LHC ring. The polar angle $\theta$ is measured with respect to the $z$ axis and the pseudorapidity $\eta$ is defined as $-\ln[\tan(\theta/2)]$.

Low-mass particles typically imply production peaking in the forward direction.
However, even decay products of a 125 GeV Higgs
boson ($H(125)$) can have small enough polar angle $\theta$ to reach FACET~\cite{Boiarska:2019vid}.
Small couplings often imply long lifetimes, hence the focus on searches for LLPs that can manifest as displaced vertices,
e.g., in the large central detectors at the LHC. Fixed-target experiments, such as NA62~\cite{NA62:2017rwk,Drewes:2018gkc} at the CERN SPS, have sensitivity for LLPs with mass less than 1 GeV from $\pi^0, \; \eta$, and $\eta'$ meson decays.

Longer lifetimes can be probed in the LHC Run 3 by the FASER experiment~\cite{Feng:2017uoz,FASER:2018bac} in the beam direction ($\theta = 0^\circ$) at a distance $z = 480$ m from the ATLAS interaction point (IP1) and by the SND@LHC~\cite{Ahdida:2750060} experiment located on the opposite side of IP1, as well as in a more central region by the MilliQan experiment~\cite{milliQan:2021lne}, located near the CMS detector and MoEDAL-MALL/MAPP experiment~\cite{Pinfold:2019zwp,Pinfold:2019nqj} located at the LHCb interaction point IP8. There are several proposed high-luminosity LHC (HL-LHC) era experiments targeting LLP searches, such as MATHUSLA~\cite{MATHUSLA:2018bqv,Curtin:2018mvb}, CODEX-b~\cite{Aielli:2019ivi}, FASER-2~\cite{FASER:2018eoc,FASER:2019aik}, ANUBIS~\cite{Bauer:2019vqk}, FORMOSA~\cite{Foroughi-Abari:2020qar}; as well as  SHADOWS~\cite{Baldini:2021hfw}, a beam-dump experiment at the CERN SPS. There are also experimental proposals outside CERN, such as SUBMET~\cite{Kim:2021eix} at J-PARC and LUXE~\cite{Abramowicz:2021zja} at the DESY XFEL facility. 

FACET is not proposed to be a new experiment, but a new subsystem of the CMS experiment that,
while overlapping in the parameter space with other searches, will cover an extended and unique region. FACET will be sensitive to particles produced with 
polar angles $1 < \theta < 4$~mrad (equivalently $7.6 > \eta > 6.2$). It is
closer to IP5 than FASER (FASER-2) is to IP1, with four hundred (four) times the solid angle. Conceptually, the FACET detector location is somewhat similar to that of the FASER ``near detector" option discussed in ref.~\cite{Feng:2017uoz} and later dropped from consideration. FACET differs in many aspects from the FASER near detector idea, as it is conceptually designed to cope with large backgrounds and radiation exposure at low polar angles with respect to the beamline, which rendered the FASER ``near detector" not a viable option.

FACET has an 18~m long decay volume from $z = 101$ to 119~m, followed by an 8~m long region instrumented with various particle detectors.
FACET covers a range of proper lifetimes $c \tau$ of $\sim$0.1--100~m. We note that the Lorentz factor $\gamma$ is typically high in the forward direction.
A unique feature among the LHC experiments is that the decay volume is at high vacuum (LHC quality, as it is part of the LHC beam pipe), eliminating any background from particle interactions inside a $\sim$14~m$^3$ fiducial region.

Small couplings also imply the ability of an LLP to penetrate a large amount of absorbing material. Between IP5 and the decay volume
LLPs have to penetrate 35--50 m (200--300 $\lambda_{\rm int}$) of magnetized iron in the LHC quadrupole magnets Q1--Q3 and the new (for Run 4) 35~T$\cdot$m superconducting dipole D1. Since neutrinos are the only SM particles that can penetrate that much
absorber, essentially all the SM backgrounds having \emph{direct} paths from the IP are eliminated. Nevertheless, the detectors are in a region with high radiation levels and particle showers from upstream interactions in the beam pipe and surrounding material. The design of FACET takes these challenges into account. 

\section{FACET as a New Subsystem of CMS}

\begin{figure}[hbt]
 \begin{center}
\makebox[0.95\textwidth][c]{\includegraphics[angle=0,origin=c,width=0.95\textwidth]{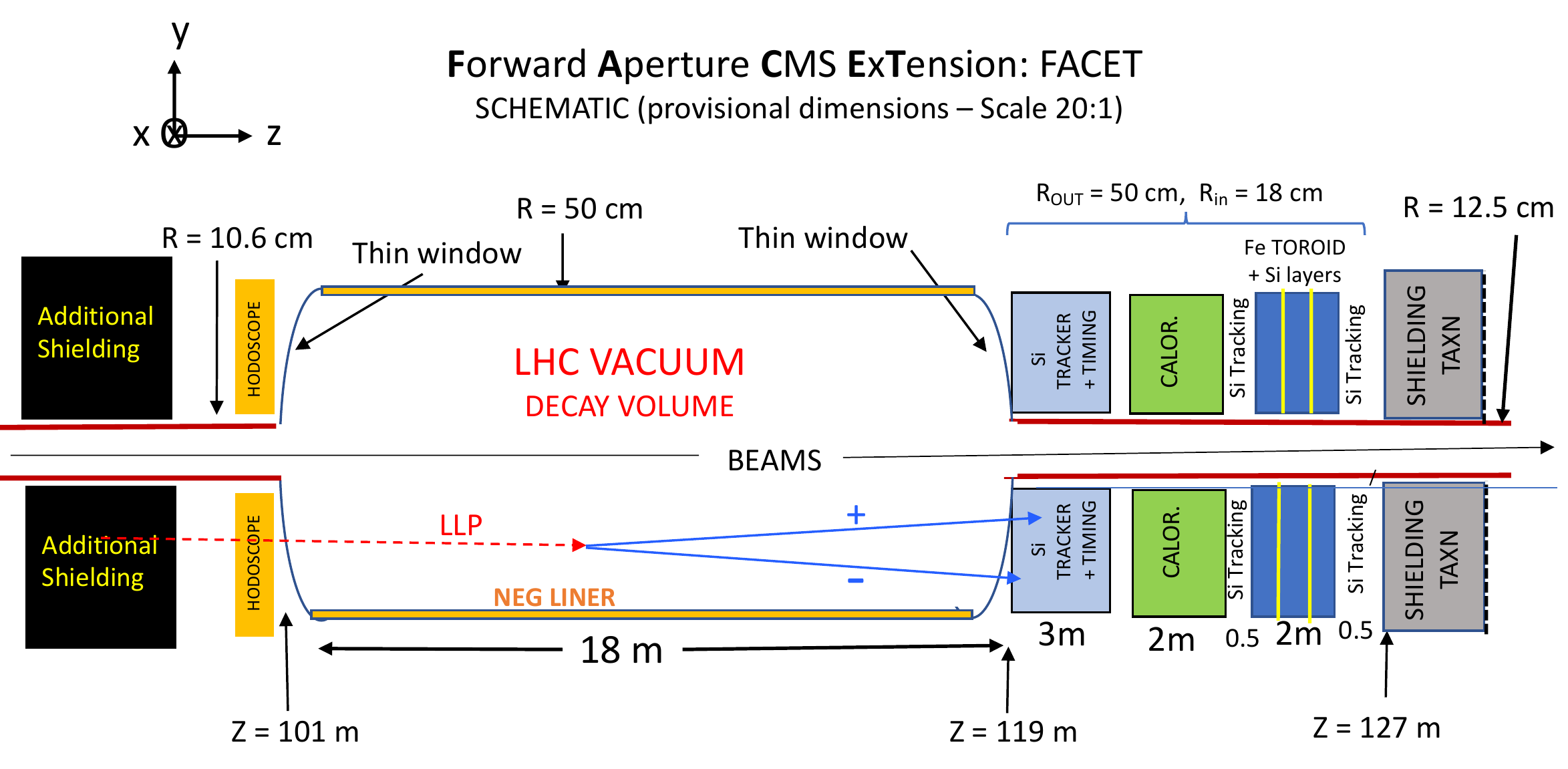}}
\vspace{-0.2in}
\caption{Schematic layout of the proposed FACET spectrometer. Side view and top view are
essentially the same since it is azimuthally symmetric. Upstream (to the left) is the IP5 collision region (at $z = 0$) followed by the machine elements Q1--Q3 and D1
comprising 35--50~m of iron shielding.
The superimposed red dashed line shows schematically an LLP from the IP5 decaying into two charged tracks shown in blue.
The TAXN copper absorber is conceived to intercept the neutral particles produced at the interaction point at the HL-LHC, in order to shield the superconducting magnets downstream. Its location limits the extent of FACET in the $z$ direction.
\label{sketch}}
\vspace{-0.2in}
\end{center}
\end{figure}

A schematic view of the detector is shown in Fig.~\ref{sketch}. The project
requires that an 18~m long section of the LHC beam pipe, between $z  = 101$ and 119~m on  one  side  of  the  
IP5  interaction  region  be  replaced  with  a  circular pipe  of a 50~cm radius\footnote{A large beam pipe with a similar radius already exists downstream of the ALICE experiment.}. This section is downstream of the focusing quadrupole magnets and beam separation dipole magnet D1. Additional shielding will be placed upstream of the first detector, which is a two-layer counter hodoscope made of radiation-hard quartz pads. The hodoscope must have very high efficiency to tag charged particles from interactions in the upstream shielding, most of which have large enough polar angles to miss the tracker. A preliminary design has  about 1500 quartz pads of size $2 \times 2$~cm$^2$ arranged in two overlapping planes to avoid cracks. The occupancy per bunch crossing will be $\lesssim$10\%, with the efficiency for charged particles $>$99.99\%.

Dedicated \textsc{fluka}~\cite{fluka:2015,BATTISTONI201510} simulation predicts that with the present design there will be $\sim$30 charged particles in the tracker per bunch crossing. These are nearly all background tracks, which are detected in the hodoscope or seen to originate from the upstream beam pipe and ignored in the subsequent analysis. The \textsc{fluka} simulation also predicts that there will  be, on average, one neutral hadron (mostly $K^0$ or $\Lambda$) decaying inside the decay volume. All bunch crossings will be examined for candidate decays inside the vacuum volume, giving a sensitivity to an integrated luminosity of $\sim$3~ab$^{-1}$ in the high-luminosity LHC era. The highly segmented upstream hodoscope, precision tracking, and imaging calorimetry will reduce these backgrounds to very low levels, even eliminating them in some channels, as discussed below. In addition, we are investigating the possibility to further mitigate these backgrounds by installing additional shielding closer to the D1 dipole and possibly shortening the vacuum decay volume by 1--2~m to make space for more shielding, and possibly a magnetized iron toroid, in front of the hodoscope.  
 
On both ends of the enlarged beam pipe, where it transitions from $R = 10.6$ to 50~cm (front) and from 50 to 12.5~cm (back), are thin ($\sim$2~mm) aluminum convex caps to minimize multiple scattering of charged particles. The convex windows also mitigate the impedance mismatch.
Behind the downstream cap (in air) precision tracking, high-granularity electromagnetic and hadron calorimetry, and a toroidal magnet interspersed with tracking detectors measure charged-particle tracks and identify and measure the energies of photons, electrons, hadrons, and muons. There is no magnetic field between the $X^0$ decay and the precision tracker, making accurate reconstruction of decay vertices simple. A layer  of  fast  timing   with  low-gain  avalanche  detectors will be included. 
This high-resolution timing, with $\sigma_t \sim 30$~ps, will provide vertex positioning in 4D ($x,y,z,t$) and a time-of-flight measurement for charged particles. These will be used to reject backgrounds from pairs of tracks from independent collisions within the same bunch crossing, which have a time spread of $\sim$200~ps. While this spread limits our ability to perform the mass measurement using time-of-flight plus energy information for the LLP candidates, this information still provides an added constraint. For example, a particle with a mass $m_{\rm X} = 5$ GeV and momentum $p = 100$~GeV has a time-of-flight from the IP5, which is $\sim$500~ps longer than for a particle with $\beta = 1$.

The  tracking  is followed  by  electromagnetic  and  hadron calorimetry,  using  identical technology 
to the CMS HGCAL (High-Granularity Calorimeter)~\cite{HGCAL} planned for the forward direction in the CMS Phase-2 upgrade.  Copper or tungsten plates interspersed with silicon pads provide imaging in 4D. The high granularity is important to measure individual showers above a threshold energy (e.g., 10 GeV, but tunable) and their directions in the presence of many low-energy showers. Behind the calorimeter, an iron toroid with magnetic field of $B \sim 1.75$~T instrumented with additional silicon tracking measures the charge of  muons  and  allows  an  approximate  measurement  of the muon momentum and the dimuon mass  for  any  muon pairs. Muons  are  also  detected through  the  active  layers  of  the calorimeter.

A dedicated \textsc{Geant4}~\cite{Geant4} model of the FACET detector is being built, which would allow us to study the backgrounds in much more detail than in the present initial paper. Nevertheless, most of the detector performance characteristics have been already obtained using fast parametric simulation. In particular, for LLPs decaying into a pair of charged particles, we expect the decay vertex position resolution in the $z$ direction from the tracking (5 planes of 100~$\mu$m silicon strip sensors with a typical hit position resolution of 30~$\mu$m)  to be $\sim$5~cm, which would allow ensuring the vertex to be located within the decay pipe without losing much of its fiducial volume. For photons, the pointing resolution in $\theta$ achievable with the HGCAL technology is $\sigma_\theta/\theta \sim 2.5\%$~\cite{HGCAL}, which translates into the decay vertex $z$ position resolution of about 15~cm. Moreover, by combining the energy information from the calorimeter and pointing information from the tracker, one could achieve a mass resolution of $\sim$5\% for LLPs decaying to a pair of hadrons or a pair of electrons. Similar, but slightly worse mass resolution ($\sim$8\%)  can be also achieved for the dimuon decay by combining the information from the tracker and the muon toroidal detector. This level of performance would provide excellent handles in suppressing various backgrounds, discussed in more detail in section~\ref{sec:backgrounds}.

The approximate number of channels in the FACET detector amounts to about 5\% of that for the CMS Phase-2 upgrade, making the detector relatively inexpensive, as most of it could be built using the same modules as will be used for the central CMS detector upgrade, thus minimizing the R\&D and engineering needs. Based on detailed \textsc{fluka} simulations, the radiation dose for the inner parts of the FACET detector is expected to be similar to that in the central CMS, so the choice of the CMS radiation-hard technology would allow the detector to withstand harsh radiation conditions at the HL-LHC without the necessity to replace the detector components, perhaps with the exception of the inner part of the forward hodoscope, which is easily accessible and could be replaced. While \textsc{fluka} simulations are a standard tool to estimate the radiation doses at the LHC, we also plan to measure the radiation background with dedicated sensors near the proposed FACET location during the forthcoming LHC Run 3, provided that we obtain the LHC approval for these measurements.

\section{Sensitivity to Long-Lived Particles}

The reach in LLP parameter space has been calculated for dark photons, heavy neutral leptons, axion-like particles, and dark Higgs bosons in several benchmark scenarios. Predictions are generally model dependent and some also depend on the nature of other BSM particles, e.g., a heavy $Z'$ boson and its mass. 
We base these studies on a total integrated luminosity of 3~ab$^{-1}$ of proton-proton collisions at a center-of-mass energy  $\sqrt{s} = 14$~TeV, with either 3 or 5 candidate events, assuming no background and that FACET can detect all penetrating neutral particle decays to $\geq 2$ charged particles or photons occurring between $101 < z <  119$~m with the decay products within $18 < R < 50$~cm at $z = 120$~m. The typical momentum spectrum of LLPs coming from meson decays peaks at 100--250~GeV, with the median in the 200--500~GeV range, depending on the LLP mass.

\subsection{Dark Photons}

Massive dark photons $A'$  are neutral gauge bosons,  
which are not directly charged under SM gauge groups. However, they  can  interact with SM particles via mixing with photons. A recent review can be found in ref.~\cite{Caputo:2021eaa}.  
A  massive  virtual  photon  produced  by  any  process  in a hadron-hadron collision has some probability 
of conversion to an $A'$, governed by the kinetic mixing parameter $\epsilon$.  If $m_{A'} \lesssim  1$~GeV, the most prolific source will be decays of $\pi^0$, $\eta$, and $\eta'$ mesons.
 The fluxes of these 
particles are highest at small polar angles.

\begin{figure}[htb]
 \begin{center}
\makebox[\textwidth][c]{\includegraphics[angle=0,origin=c,width=4.5in]{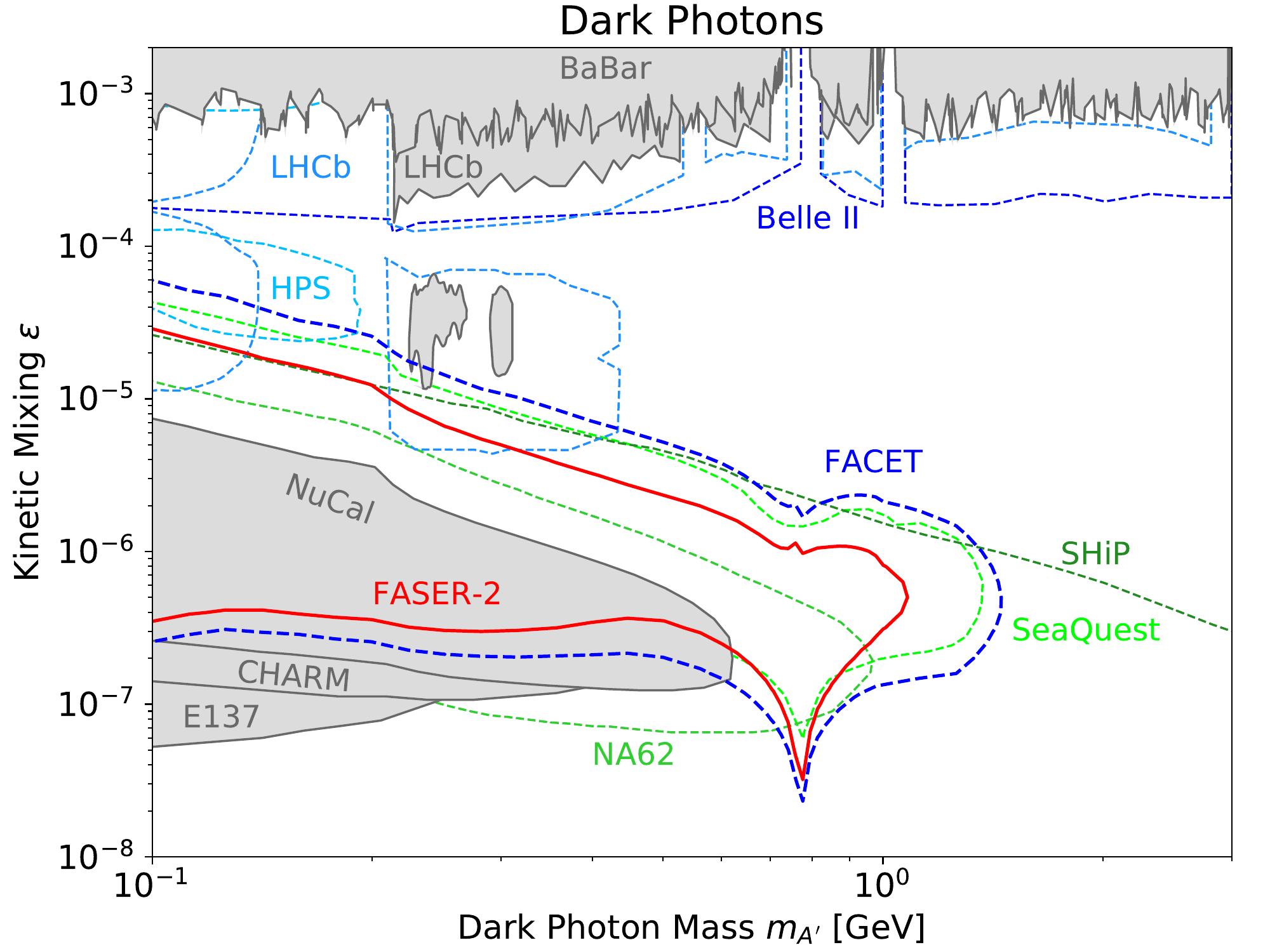}}
\vspace{-0.2in}
\caption{FACET reach for dark photons (3 event contours) in a generic model with no BSM sources, as calculated with \textsc{Foresee}~\protect\cite{Kling:2021fwx}. Existing bounds (gray shaded regions) are taken from CHARM (following Ref.~\cite{BABAR:PhysRevLett.113.201801}), 
BaBar~\cite{BABAR:PhysRevLett.113.201801}, 
E137~\cite{E137:PhysRevD.38.3375}, 
LHCb~\cite{LHCb1:PhysRevLett.124.041801}, 
NuCal~\cite{Blumlein:2013cua,Blumlein:2011mv}, along with the prospective limits taken from studies performed for 
Belle~II~\cite{Belle:2020the}, HPS~\cite{HPS1, HPS2}, 
LHCb~\cite{LHCb2:PhysRevD.92.115017, LHCb3:PhysRevLett.116.251803}, 
NA62~\cite{NA62:Dobrich:2018ezn}, 
SeaQuest~\cite{SeaQuest:PhysRevD.98.035011}, 
FASER-2~\cite{Kling:2021fwx}, and 
SHiP~\cite{SHiP:ahdida_sensitivity_2021}.
\label{Aprimeforesee}
}
\vspace{-0.2in}
\end{center}
\end{figure}

Fig.~\ref{Aprimeforesee} shows limits calculated using the \textsc{Foresee} package~\protect\cite{Kling:2021fwx} without assuming any other BSM sources of dark photons, such as heavy $Z'$ bosons, which can extend the mass range and require the energy of the LHC. 

For  $m_{A'} >  1$~GeV  the  main production mechanisms  are: $q + \bar{q} \to A' + X$; Drell--Yan: $q + \bar{q} \to A'$; bremsstrahlung: $p \to A' + p$ and $q \to A' + q$; and
heavy-quark decays: $c \to A' + X, \: b \to A' + X$. The decay modes to SM particles of a minimal dark photon are the same as the final states in $e^+ e^- \to \gamma^*$ at $\sqrt{s} = m_{A'}$.

\begin{figure}[htb]
 \begin{center}
\makebox[\textwidth][c]{\includegraphics[angle=0,origin=c,width=3.26in]{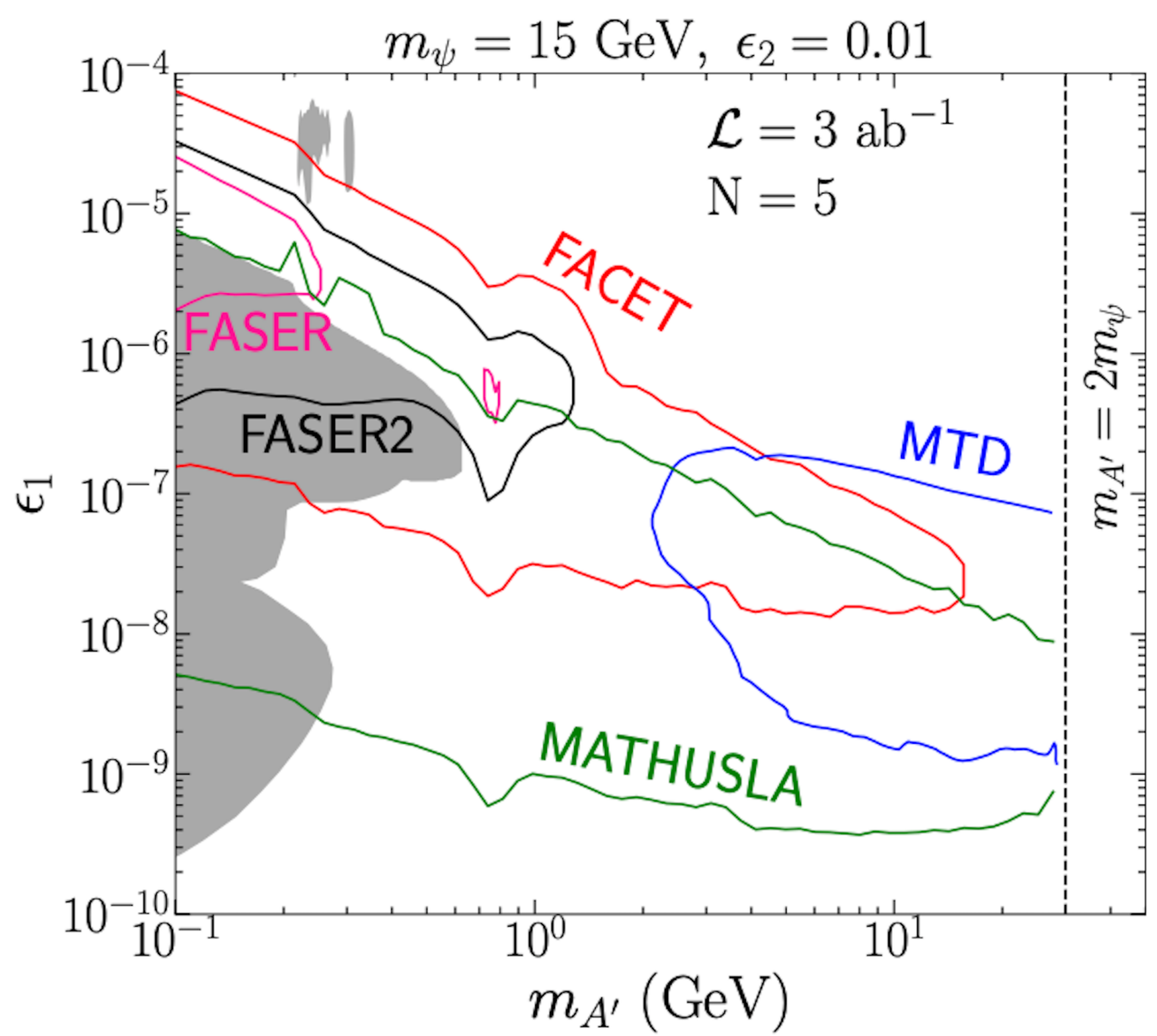}
\includegraphics[angle=0,origin=c,width=3.2in]{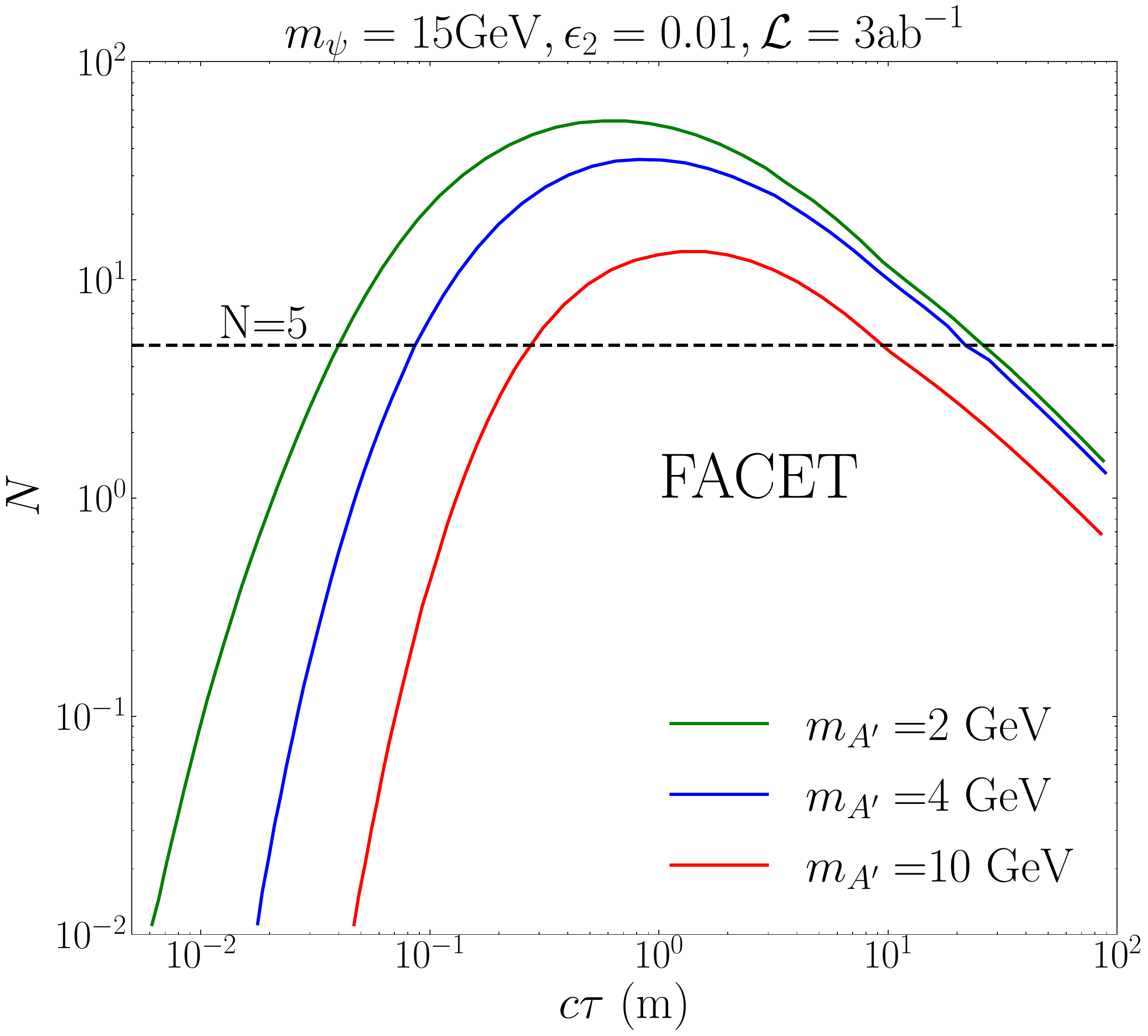}}
\vspace{-0.2in}
\caption{Left: FACET reach for dark photons (5 event contours) 
in the parameter space of coupling $\epsilon_1$ and mass $m_{A'}$ in the model of 
ref.~\cite{du2021enhanced}. Of the other projects shown, only FASER and MTD, the CMS Phase-2 MIP Timing Detector, are currently approved. Right: Number of dark photon events as a function of $c \tau$ for $m_{A'}$ = 2, 4, and 10~GeV in this model.
\label{Aprimeem}}
\vspace{-0.2in}
\end{center}
 \end{figure}

A comparison of the reach of FACET and other experiments for dark photons in all final states in the model of ref.~\cite{du2021enhanced} is given in figure~\ref{Aprimeem} (left). In this model, the main production mechanism of dark photons is via radiation in a rather rich hidden sector, which contains a Dirac fermion $\psi$ and two gauge bosons, which mix with the SM weak hypercharge field $B_\mu$. FACET covers a unique region of the mixing parameter $\epsilon_1$ vs. mass (or alternatively lifetime vs. mass) phase space. Figure~\ref{Aprimeem} (right) shows the number of events as a function of lifetime $c\tau$ for three $A'$ masses for the model parameters corresponding to the reach shown in figure~\ref{Aprimeem} (left). 

Our reach estimates do not include dark photons that are created in hadronic showers, e.g., from $\pi^0$ and $\eta$ meson decays in the iron absorber, and are therefore conservative.

\subsection{Heavy Neutral Leptons}

Many extensions of the SM involve heavy right-handed neutrinos or heavy neutral leptons $N_i$ (where the subscript $i$ indicates flavor),
which may 
explain the light neutrino masses through the seesaw 
mechanism~\cite{Minkowski:1977sc,Chang:1984qr,SHiP:2018xqw}. 

These particles may  be  produced  in  any  kinematically allowed SM weak 
leptonic decay, e.g.,  of $s$, $c$, $b$, $t$ quarks, or $W$  or $Z$ bosons. We consider a specific extension of the SM~\cite{Deppisch:2019kvs}, with a $Z'$ boson (which can be light and  yet  have  escaped  detection  due to the small coupling to SM particles) and three heavy 
right-handed Majorana neutrinos $N_i$. In this model, the decay $Z' \to N_i N_i$ is allowed, and the $N_i$ can be
long-lived  and  decay  to  SM  particles, e.g., a lepton of the same flavor and a virtual $W^*$ or $Z^*$ boson.  For the  $Z'$ masses in the 10--100~GeV range, most interesting in this model, the branching fraction of the $Z' \to N_i N_i$ decays amounts to about 20\%, i.e., rather large and similar to that for the $Z$ boson decays into SM neutrinos.

 \begin{figure}[htb]
\vspace{-0.25in}
 \begin{center}
\makebox[\textwidth][c]{\includegraphics[angle=0,origin=c,width=3.3in]{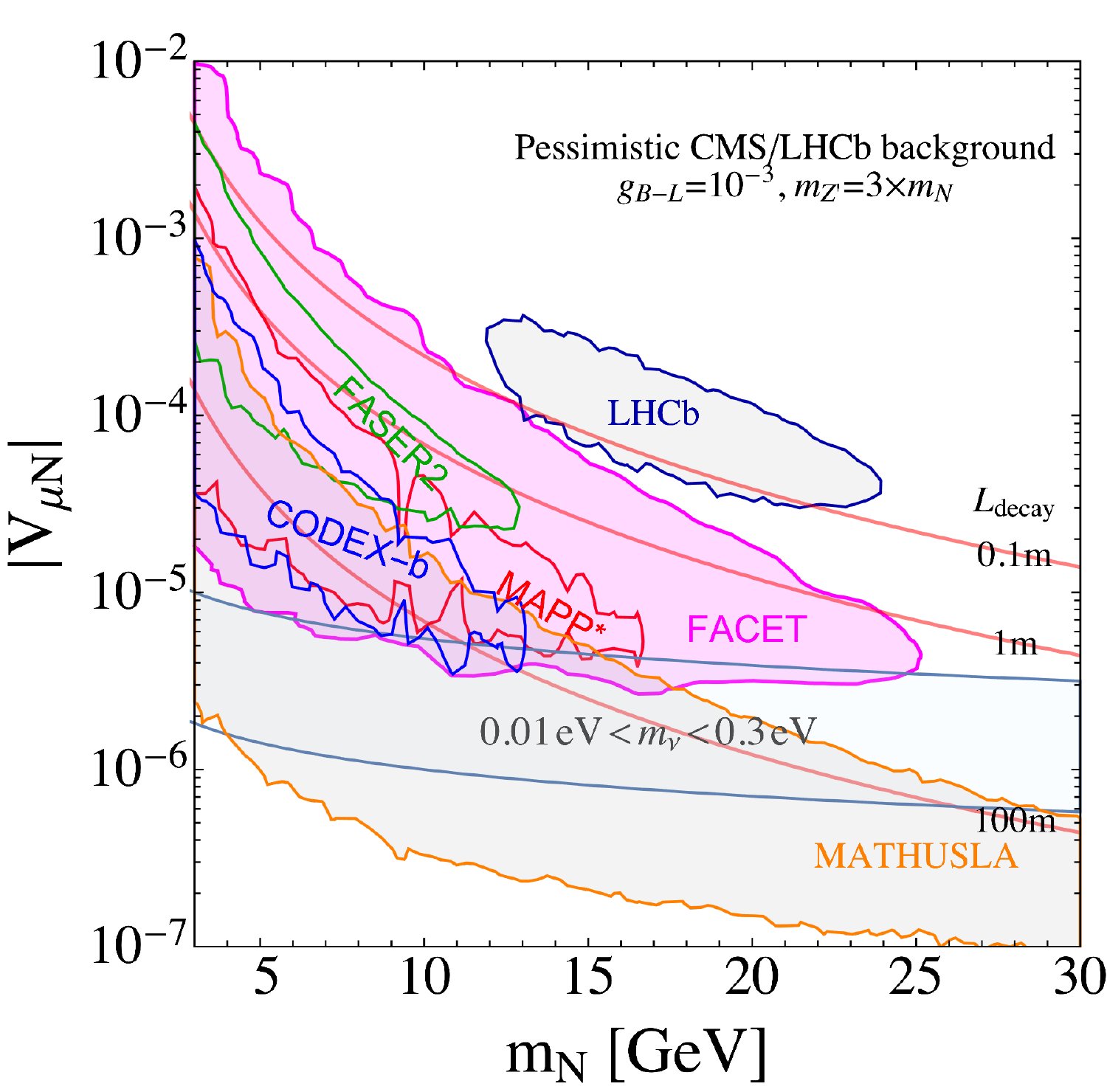}}
\vspace{-0.2in}
\caption{FACET reach in the mixing parameter vs. mass plane for a heavy neutral lepton (3 event contours), along with projections for other proposed experiments, as well as for MAPP and the upgraded LHCb detectors~\protect\cite{Deppisch:2019kvs}.
\label{hnlsuchita}}
\end{center}
\vspace{-0.2in}
 \end{figure}
 \begin{figure}[!htbp]
 \begin{center}
\makebox[\textwidth][c]{\includegraphics[angle=0,origin=c,width=4.4in]{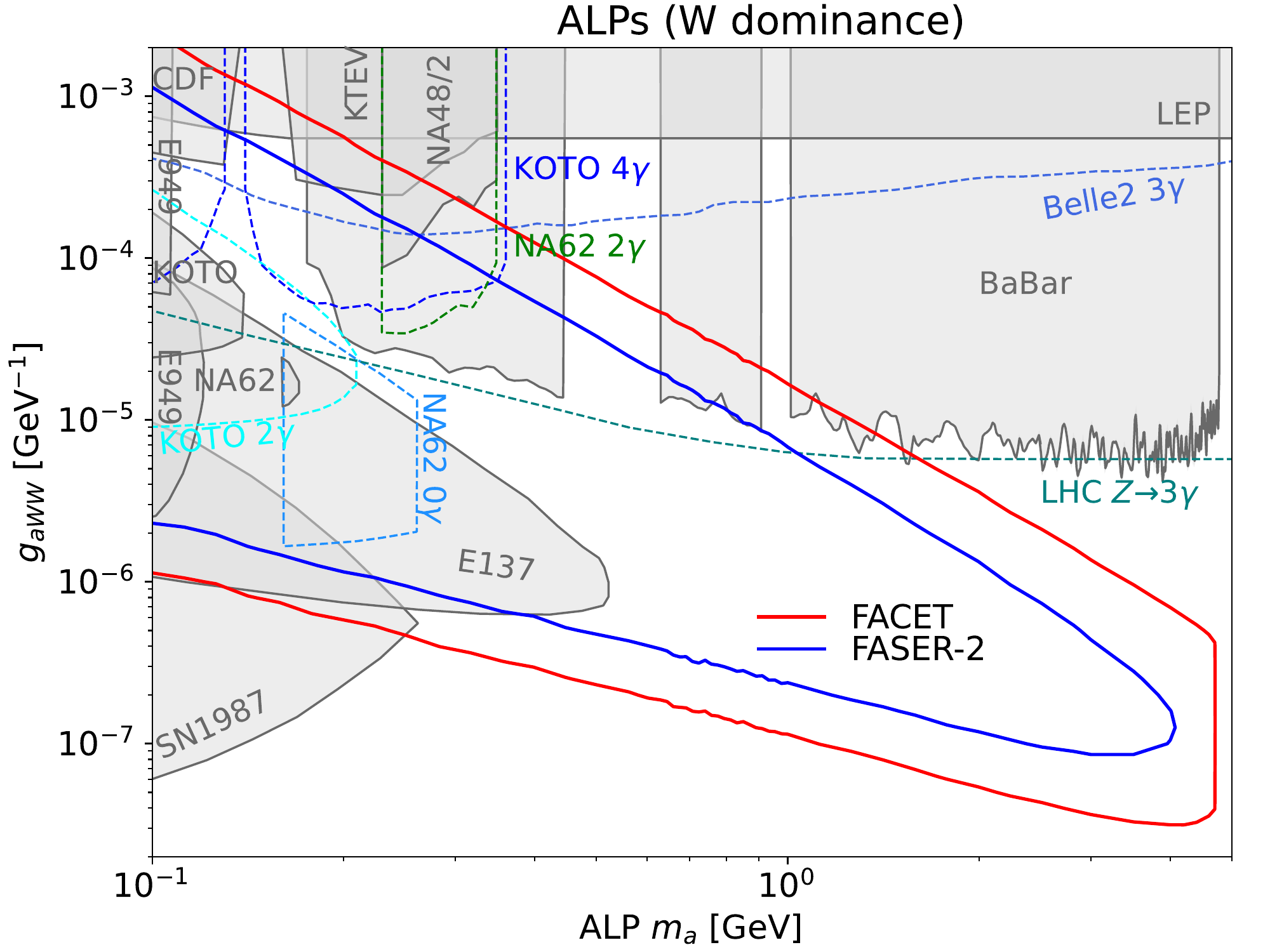}}
 \vspace{-0.2in}
\caption{FACET sensitivity to ALPs in the $W$-dominance model (3 event contours), as a function of mass and coupling, as calculated with \textsc{Foresee}~\protect\cite{Kling:2021fwx}. The gray-shaded regions are excluded by current bounds, while dashed lines correspond to projected sensitivity of various experiments, as calculated in refs.~\cite{Gori:2020xvq,Kling:2020mch}. The BaBar limits are from ref.~\cite{BaBar:2021ich}.
More details can be found in refs.~\cite{ALPs-1:Beacham_2019, ALPs-2:dolan_revised_2017} and references therein.
\label{alps}}
\vspace{-0.3in}
\end{center}
\end{figure}

Fig.~\ref{hnlsuchita} shows the coverage in the mixing parameter $|V_{\mu N}|$ vs.  $m_N$ plane in the case of a single Majorana neutrino $N$ mixed with a muon neutrino.  
In this case, FACET has a unique sensitivity at high masses, above $\sim$15~GeV for lifetimes $c\tau$ between $\sim$0.1 and $\sim$100~m.

\subsection{Axion-Like Particles}

Pseudoscalar particles, such as extremely light axions,
were initially proposed to  solve the strong-$CP$ problem of QCD. 
More massive axion-like particles (ALPs, $a$) may
exist, and if produced at the LHC~\cite{dEnterria:2021ljz,mimasu2015alps},
they may decay  with  long  
lifetimes  into  photon  pairs  (or $\gamma e^+e^-$) or  lepton  pairs,  after  penetrating  thick  absorbers. FACET will be well-placed to 
discover such ALPs in certain regions of their mass and the coupling to SM gauge bosons. An overview of the FACET reach for
ALPs is given in figure~\ref{alps} in a specific $W$-dominance ALP model~\cite{Gori:2020xvq,Kling:2020mch}, as a function of the ALP mass and the coupling to $W$ bosons, $g_{aWW}$.

\begin{figure}[!htbp]
\begin{center}
\makebox[\textwidth][c]{\includegraphics[angle=0,origin=c,width=3.24 in]{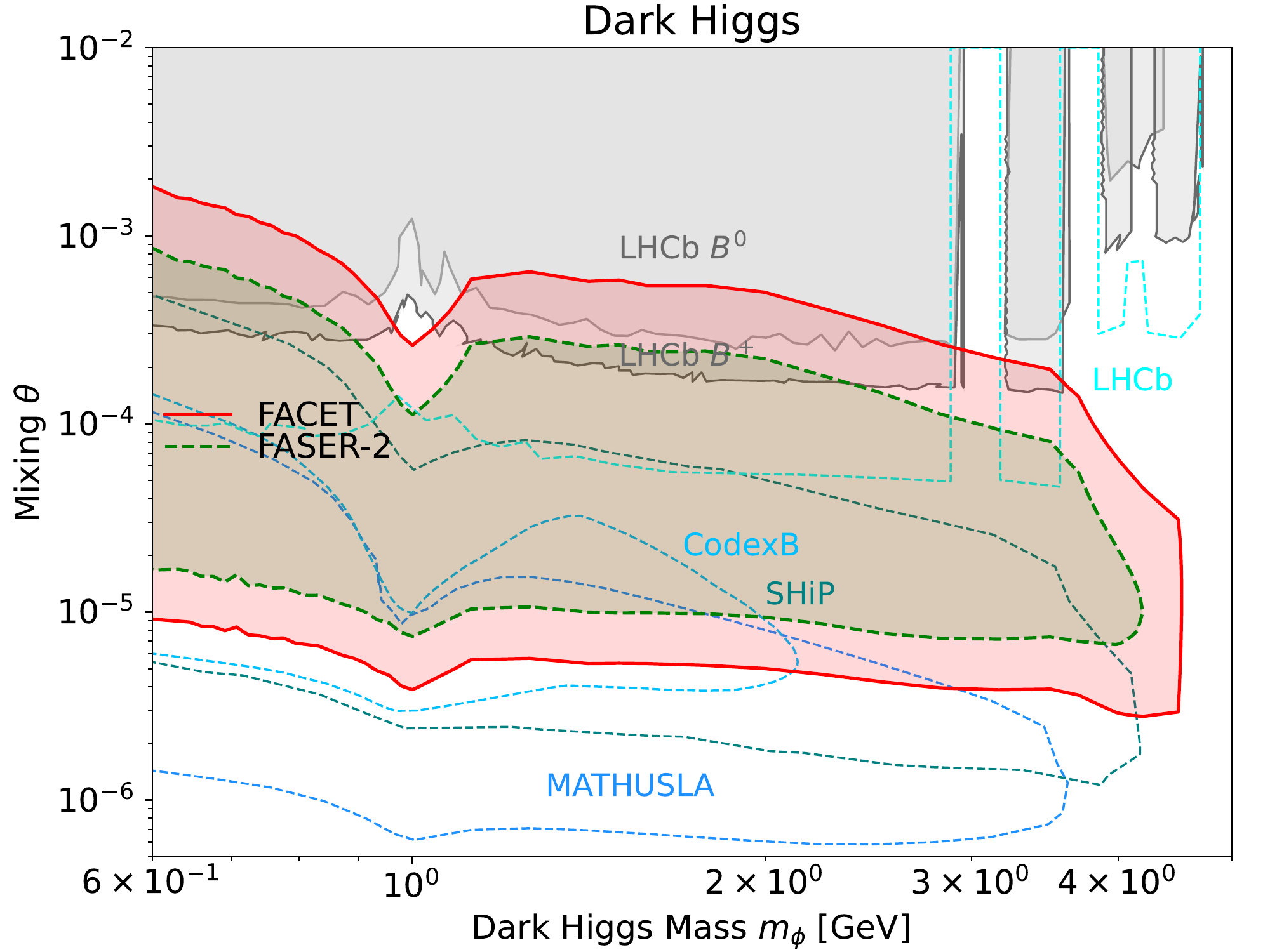}
\includegraphics[angle=0,origin=c,width=3.24 in]{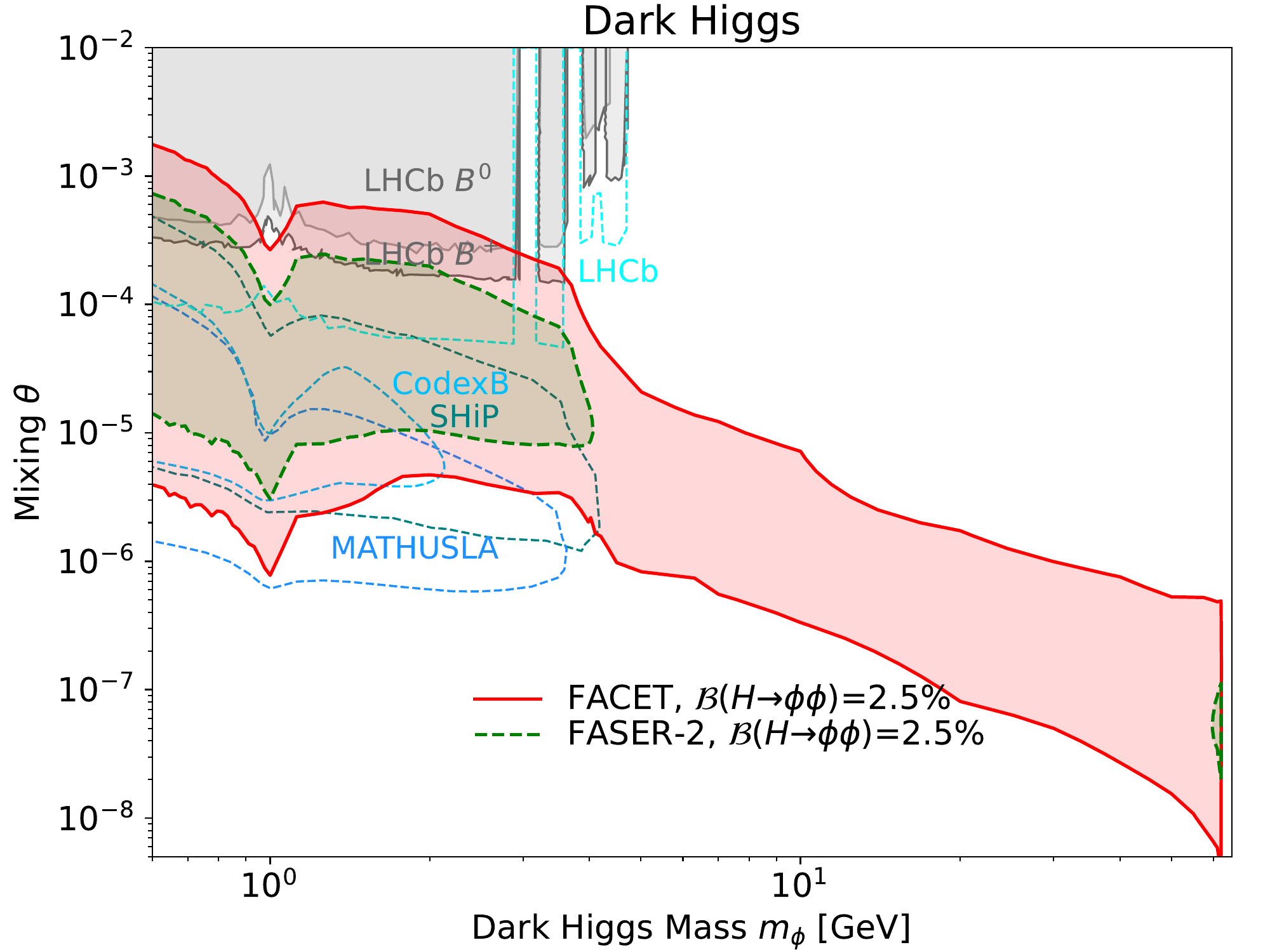}}
\vspace{-0.2in}
\caption{Reach of FACET and other existing and proposed experiments for a dark Higgs boson $\phi$ (3 event contours)
with the assumption of either 0\% (left) or 2.5\% (right) branching fraction for the  $H(125) \to \phi\phi$ decays. In the second scenario, FACET offers a unique coverage all the way to half $m_H$ for a range of mixing angles. FACET and FASER-2 contours are calculated with \textsc{Foresee}~\protect\cite{Kling:2021fwx}.
Current exclusions from LHCb \cite{LHCb:2015nkv,LHCb:2016awg} are shown as gray shaded regions. Also, sensitivity from studies for 
MATHUSLA~\cite{MATHUSLA-DarkHiggs:CHOU201729}, CODEX-b~\cite{CodexB-DarkHiggs:PhysRevD.97.015023}, LHCb upgrade~\cite{CodexB-DarkHiggs:PhysRevD.97.015023}, and
SHiP~\cite{SHiP-DarkHiggs:Alekhin_2016} are shown. 
\label{darkhiggs}}
\vspace{-0.2in}
\end{center}
\end{figure}

\subsection{Dark Higgs Bosons}

The possible existence of a dark-sector partner $\phi$ of the 125~GeV Higgs boson has attracted 
attention, as discussed, e.g., in refs.~\cite{Duerr:2017uap,Feng:2017vli,Bertuzzo:2020rzo}. A dark Higgs field provides a simple mechanism to give mass to the dark photon $A'$. 
 The corresponding dark Higgs boson $\phi$ may be the lightest dark sector state and can decay into SM particles via mixing with the Higgs boson, governed by the mixing angle $\theta$.  Unitarity and perturbativity suggest that the dark Higgs boson cannot be much 
heavier than the dark gauge boson $A'$, while it can be significantly  lighter. The dark Higgs boson can be very 
long-lived due to its suppressed couplings to the accessible light SM states. 

For mass ranges below $\sim$5~GeV the dominant production 
mechanism is through $B$ meson decays, e.g., $B \to K + \phi$, with $\phi$ decaying to pairs of most massive SM fermions accessible kinematically, e.g., to a pair of muons for light $\phi$, or to a pair of $\tau$ leptons or charm quarks for a heavier $\phi$. A heavier $\phi$ may also decay to another pair of new scalars, $s$, which may in turn be LLPs. 

The reach of FACET for a dark Higgs boson decaying to a detectable final state is given in figure~\ref{darkhiggs}. In addition to the production via $B$ meson decays (left plot), we also consider the case of a small, non-zero trilinear coupling $\phi\phi H$ between the SM Higgs and dark Higgs bosons, resulting in a 2.5\% branching fraction of the $H(125) \to \phi\phi$ decay (right plot). This value is lower than the projected limits on the BSM Higgs boson decay branching fraction at the HL-LHC~\cite{Dainese:2019rgk}. In this case, the low-mass reach is slightly improved compared to the case with no trilinear coupling, as a new decay mode $b \to s\phi\phi$, where $b$ and $s$ are the bottom and strange quarks, respectively, is present due to a virtual Higgs boson exchange~\cite{Feng:2017vli}. However, the most striking feature in this model is that FACET offers a unique sensitivity for the dark Higgs boson masses up to the kinematic limit of $m_H/2$ due to a large number of Higgs bosons that are produced at the HL-LHC with a significant longitudinal boost, resulting in at least one of the two dark Higgs bosons decaying within the FACET detector acceptance~\cite{Boiarska:2019vid}.

\section{Triggers}
As a part of the CMS Phase-2 Upgrade at the HL-LHC, the Level-1 (L1) trigger system will be upgraded, with a latency increased to 12.5~$\mu$s
and output rate to the High-Level Trigger (HLT) increased to 750~kHz. The HLT will analyse the data
with close to off-line performance, sending data to long-term storage at a rate of about 7.5~kHz.
FACET will provide an additional external trigger to the CMS L1 Global Trigger, 
 built from the hodoscope, tracking,  calorimeter, and muon detector information, and using the same
hardware and code to be used in the upgraded CMS detectors.

The L1 triggers will  be formed from, among others:
\begin{enumerate}
\item $\ge \;$2 tracks with a small distance of closest approach inside a decay volume;
\item $\ge \;$2 muon tracks through the toroidal spectrometer;
\item $\ge \;$1 cluster of energy in the electromagnetic calorimeter above some threshold and with a direction requirement based on the tracker and/or the calorimeter;
\item $\ge \;$1 cluster of energy in the hadron calorimeter above some threshold and with a direction requirement based on the tracker.
\end{enumerate}

To achieve maximum sensitivity for the LLP search, FACET will be exposed to all bunch crossings. The goal of the trigger is to select all candidates for $X^0 \to \; \ge$ 2 charged tracks or two photons (even if merged), while excluding decays of the SM particles, such as $K^0$ and $\Lambda$.

The \textsc{fluka}~\cite{BATTISTONI201510, fluka:2015} code, regularly used for LHC background calculations,
predicts that
there will be about 30 charged particles with the momentum above 1 GeV (mostly protons, $\pi^\pm$, and $e^\pm$, with 
$\sim$1.9 $\mu^\pm$) 
entering the tracker at $R > 18$~cm per bunch crossing (with a pileup of 140). 
These are all background tracks, and
will be tagged as such by the front hodoscope and ignored. The L1 track trigger will form
tracks (at $119 < z < 122$ m) and calculate the position of candidate vertices inside the decay volume, as well as confirm that the decay signature is consistent with an LLP originating at IP5. 

Since the planned rate of L1 triggers for CMS is 750 kHz, FACET triggers at a L1 rate of a few kHz is a goal, which should be achievable by tuning thresholds. The HLT can apply selections close to the offline analysis to reduce the rate to long-term storage to $\lesssim$100~Hz out
of 7.5 KHz total rate-to-tape for CMS. The FACET-triggered events will include the full CMS data, and the FACET information
will be included in all CMS triggered events. The FACET data will be $\ll 1\%$ of
the full CMS data. The FACET trigger could also be run in a standalone mode, with only FACET information saved,
and without correlating with the central CMS detector.

For charged particles with $\theta < 1$~mrad that come through the dipole D1 aperture, the rates will be high, but some SM channels, e.g.,  $e^+ e^-$ and $\mu^+\mu^-$, are interesting and special triggers for those can be included.
Such triggers will be prescaled, but the data will be useful for checks throughout the LHC running.

For any  low-pileup LHC runs, with proton, as well as with ion beams, a different set of triggers will be prepared.
Since many bunch crossings will then have only a single interaction, correlations between leading charged hadrons and the central event can be studied.

\section{Backgrounds}
\label{sec:backgrounds}

FACET is unique among
all LHC LLP search proposals in having a very large volume of LHC-quality vacuum for decays. Vertices with $R > 15$~cm inside a fiducial volume with two or more associated tracks cannot have come from 
interactions; they must be due to decays\footnote{Interactions of beam bunches with residual gas molecules in the LHC during the HL-LHC operations are actively being
studied \cite{Bruce:2019rpx}, but such vertices would be close to the outgoing beams with $R < 11$ cm.}. 
Our goal is to have no background events
even with 3 ab$^{-1}$ of integrated luminosity in many decay channels; in which case even a few signal candidate events can mark a discovery.

The direct path from the collision region to the decay volume has 200--300 $\; \lambda_{\rm int}$ (depending on $\theta$) of magnetized iron, effectively eliminating 
all SM particles, except neutrinos. Therefore the only SM particles entering the decay volume
are indirect, from interactions in the beam pipe and LHC components. Most are at large enough polar angle $\theta$ to miss the tracker, nevertheless the \textsc{fluka} code predicts that
there will be about  $\sim$30 charged particles with the momentum above 1~GeV
entering the tracker at $R > 18$~cm per bunch crossing. 
There is a significantly larger flux of lower-momentum charged particles entering the hodoscope, most of which
have large enough polar angles to miss the downstream tracker. This drives the need for at least two layers of $\sim$2 $\times$ 2~cm$^2$ pads in the hodoscope.

Neutral hadrons of concern are $K^0_S$, $K^0_L$, $\Lambda$, and $\Xi^0$, with about one entering
the decay volume per bunch crossing.
Their decay tracks will be well measured and their energies
determined in the calorimeter. A Monte Carlo simulation shows that
the parent mass and the direction can be reconstructed with this information.
Requiring the parent track to point back to the IP5 interaction region 
and using the decay position information (flat in $z$ for an LLP) will reduce this neutral-hadron
background that may still be
overwhelming for $X^0 \to h^+ h^-$ (where $h^\pm$ denotes charged hadrons) with $m_{X^0} \lesssim 0.8$~GeV.

The situation is much better for lepton pairs. Only $K^0$ decays can contribute; for $K^0_S$ either through both charged pions
being misidentified as electrons or as muons, or by genuine dilepton decays,
all of which have very small branching fractions $< 10^{-8}$. The $K^0_L$ meson has common semileptonic decays to $\pi^\pm e^\mp \nu_e$ 
and $\pi^\pm \mu^\mp \nu_e$, so only one $\pi^\pm$ has to be misidentified as a lepton. 
The missing neutrino smears the pointing from IP5, the reconstructed mass is a continuum with  $m_{X^0} < 500$~MeV, and the
$z$ distribution of the vertex is not nearly uniform as it would be for an LLP.

In 2$\times$10$^{15}$ bunch crossings (3 ab$^{-1}$) we expect several thoU.S.A.nd  true $K^0_L \to \mu^+ \mu^-$ 
decays in the vacuum volume given the branching fraction of $7 \times 10^{-9}$. 
The $\mu^+\mu^-$ mass is reconstructed and, if compatible with $m_{K^0}$, the momentum, the decay time $c\tau$ and the total momentum are known. While it would
be interesting (and an excellent control measurement)
to observe these rare $K^0$ dilepton decays, they will not be a background to $X^0 \to l^+l^-$ decays
for $ m_{X^0} \gtrsim 0.6$~GeV.

A potential background in the $X^0 \to l^+l^-$ channel is from pileup, with two muons or electrons from different 
collisions in the same bunch crossing appearing to come from a common vertex in the decay volume.
Studies done with 
\textsc{fluka} show that the transverse distribution of muons is approximately proportional to $1/R$, 
the density ranging from $2 \times 10^{-4}$ to $8 \times 10^{-5}$~cm$^{-2}$. The total
is an average of 1.9 muons (both charges) per bunch crossing
within $18 < R < 50$~cm. 
This background will be eliminated by charged-particle tagging in the upstream hodoscope and precision vertexing. 
If the inefficiency of the hodoscope is $10^{-4}$ ($10^{-5}$) there will be  $\sim$10$^7$ ($\sim$10$^5$) bunch 
crossings in 3 ab$^{-1}$ with two or more untagged muons entering the decay volume.
A Monte Carlo study of pairs of uncorrelated muons was used to determine the probability that any pair has a distance of closest approach $< 60$~$\mu$m; the prediction is 150 (1.5) two-track vertices from pileup, which is further reduced by a factor of two by the opposite-sign track requirement.
Further requiring the vector sum of the muon momenta to point back to the IP5 interaction region eliminates this background.

A search for $X^0 \to \gamma \gamma$, e.g., for an axion-like particle,
having no charged tracks and less precise vertex location, will be challenging,
with a large background from photons from $\pi^0, \eta$, and $\eta'$ meson decays, as well as from $K^0_S \to \pi^0\pi^0$ decays. 
The electromagnetic section of the calorimeter measures both the shower directions and the distance of closest
approach of the two photons, albeit without the high precision which is achieved for tracks. 
Requiring matching in \emph{x,y,z,t} using position and timing information and that 
the momentum of the diphoton pair points back to IP5 will suppress these backgrounds, especially for $m_{\gamma \gamma} \gtrsim 1$~GeV. Studies based on full detector simulation are under way to determine whether this background could be controlled.

Many BSM particles with masses above about 1~GeV have decay modes to more than two charged particles.
The only SM hadrons that can decay to four charged particles inside the FACET decay volume are $K^0$, via the following decays: 
$K^0_S \to e^+e^-\pi^+\pi^- \:,
 K^0_L \to e^+e^-\pi^+\pi^-\:, 
 K^0_L \to e^+e^-e^+e^-$, and
$K^0_L \to \mu^+ \mu^- e^+e^-$. 
With the expected $K^0$ fluxes FACET will detect such decays, but they will not
constitute a background for 4-body decays with $m_{X^0} \gtrsim 0.6$~GeV.

We have also considered pileup of
two unrelated neutral-hadron (e.g., $K^0_S,\; \Lambda$) decays, but to be a background to $X^0 \to 4$ hadrons
these decays must be superimposed in \emph{x,y,z}, consistent in time,
and the apparent ``parent'' must point back to IP5. In addition, for some of the signals we may veto pair masses compatible with that of a neutral $K^0$ or $\Lambda$. These requirements eliminate the background from pileup.

To summarize, while decays of neutral hadrons inside the vacuum volume will be a major source of background for hadronic decays of LLPs with 
$m_{X^0} \lesssim 0.8$~GeV, decays to leptons and multihadrons at higher masses should have vanishing backgrounds
even in 3 ab$^{-1}$, thanks to 200--300 $\lambda_{\rm int}$ of the iron absorber, the vacuum decay volume,
high-precision tracking, a high-granularity calorimeter, and muon momentum measurement in the toroidal spectrometer. 

\section{Summary}
FACET is proposed as a new subsystem for the CMS experiment in the high-luminosity LHC era. 
The  primary  objective  is  to  search  for beyond the standard model long-lived particles 
 decaying  in  a  large  vacuum  volume,  during the  high-luminosity LHC phase, corresponding to an integrated luminosity of about 3 ab$^{-1}$ of proton-proton collisions at $\sqrt{s} = 14$~TeV.
The FACET detector requires an enlarged beam pipe section between $z = 101$ and 119~m, followed by high-precision tracking and calorimeter modules using identical technology to the CMS Phase-2 upgrade. These are designed for the high-radiation environment expected in the high-luminosity LHC era.
The searches can be background-free
in many channels, especially for neutral long-lived particles with masses $\gtrsim$1~GeV. 
FACET will  make  an  inclusive  search  for dark photons, heavy neutral leptons, axion-like particles,  and dark  Higgs bosons 
with a sensitivity defined by their masses and couplings to standard model particles. The couplings must be large enough 
to give a detectable production cross section in the forward direction, and for the particles to decay to visible states, while small enough for the particles to traverse 35--50~m of iron upstream of the decay volume.
FACET will explore a unique area in the parameter space of mass and couplings, largely complementary to other existing and proposed searches, yet with 
some overlap ensuring seamless coverage. 

\section{Acknowledgments} 
We thank V.~Kashikhin (Fermilab) for the preliminary toroid design, P.~Fessia (CERN ATS-DO) and V.~Baglin (CERN TE-VSC) for information on the LHC and beam pipe, respectively. We are grateful to Felix Kling for a number of helpful discussions related to signal simulation and the \textsc{foresee} package.
The work of G.~Landsberg is partially supported by the DOE Award No. DE-SC0010010.
S.~Kulkarni is supported by the Austrian Science Fund Elise-Richter grant project number V592-N27. 
The University of Iowa group work is supported by the DOE Award No. DE-SC0010113.
Istanbul University group work is supported by FUA-2018-32919 from the Scientific Research Projects Coordination Unit of Istanbul University.
The University of Maryland group effort is supported by the DOE Award No. DE-SC0010072. 
M.~Du, R.~Fang, and Z.~Liu are supported in part by the National Natural Science Foundation of China under Grant No.\ 11775109. 
V.Q.~Tran is supported in part by the National Natural Science Foundation of China under Grant No. 19Z103010239. 
We acknowledge support provided by the following funding agencies: Academy of Finland and HIP (Finland), TUBITAK and TENMAK (Turkey), DOE and NSF (U.S.A.).
 
\bibliographystyle{JHEP}
\bibliography{FACET}
\end{document}